\documentclass[pra,aps,superscriptaddress,showpacs,tightenlines,amsmath]{revtex4}
\usepackage{epsfig}
\usepackage{amsmath,amssymb}

\usepackage{adjustbox}
\usepackage{array}
\usepackage{booktabs}
\usepackage{multirow}

\newcolumntype{R}[2]{%
    >{\adjustbox{angle=#1,lap=\width-(#2)}\bgroup}%
    l%
    <{\egroup}%
}

\begin{document}

\title{Teleportation based on control of anisotropic Ising interaction in three dimensions}

\author{Francisco Delgado}
\email{fdelgado@itesm.mx}
\affiliation{Escuela de Ingenier\'ia y Ciencias, Tecnol\'ogico de Monterrey, M\'exico.}
\affiliation{Departamento de F\'isica y Matem\'aticas, Tecnol\'ogico de Monterrey, Campus Estado de M\'exico, Atizap\'an, Estado de M\'exico, CP. 52926, M\'exico.}

\date{\today}

\begin{abstract}

Possibly, teleportation is the most representative quantum algorithm in the public domain. Nevertheless than this quantum procedure transmits only information than objects, its coverage is still very limited and easily subject to errors. Based on a fine control of quantum resources, particularly those entangled, the research to extend its coverage and flexibility is open, in particular on matter based quantum systems. This work shows how anisotropic Ising interactions could be used as natural basis for this procedure, based on a sequence of magnetic pulses driving Ising interaction, stating results in specialized quantum gates designed for magnetic systems.



\end{abstract} 

\maketitle

\section{Introduction}
Quantum computation is possibly the up most application goal of quantum mechanics in nowadays. Feynmann devised this relation based on use of quantum properties to speed up the simulation of computational problems by physical systems \cite{feyn1, feyn2}. Thus, processing based on quantum algorithms, useful developments using quantum information processing have been effectively developed as theoretically as experimentally: quantum dense coding \cite{bennet3,wang2}, quantum key distribution \cite{ekert1,bennet5}, quantum computation \cite{nielsen1,ren1} and quantum teleportation \cite{bennet4}.

Quantum gate array computation is the most common, direct and clear approach in terms of proximity with classical computing. Their similitude is based on the use of computer gates replicated from classical programming. This gates are reproduced by several designs in terms of physical resources where they have been carried out: ion traps and electromagnetic cavities \cite{cirac1,turchette1}, Josephson junctions \cite{shnirman1}, nuclear magnetic resonance \cite{chuang1} and spins \cite{kane1,loss1,vrijen1}. Nevertheless, their translation to theoretical gates is not always immediate, requiring control or iterative procedures.

Teleportation is a physical process now in the public domain by its attractiveness. In it, a quantum state can be transferred to other using a previously shared entangled pair, with assistance of classical communications and local operations. In \cite{bennet4}, a Bell state is used to this goal together with Hadamard, $C^a NOT_b, X$ and $Z$ gates. Departing from this development, many proposals and goals have been made to transfer a multi-qubit state, inclusively in the experimental terrain \cite{bouw1,karl1,nielsen3}. Teleportation has been accomplished among optical systems \cite{bouw1,boshi1,furu1}, photons and a single atomic ensemble \cite{sher1,chen1,olm1} and trapped atomic ions using Coulomb interaction \cite{rieb1,barr1,rieb2}. Therefore, teleportation is commonly considered as one of the most striking progresses of quantum information theory.

Ising model \cite{ising1,brush1,baxter1} is used as a simple approach to magnetic interaction between quantm objects (electronic gases, quantum dots, ions, etc.). Nielsen \cite{nielsen2} was the first reporting studies of entanglement between magnetic systems based on a two spin systems driven with an external magnetic field. One property of this model is that it generates entanglement, one of the more interesting properties of quantum mechanics \cite{vneumann1, schrod1, schrod2, einstein1, schrod3}. This property is a central aspect in the most of quantum applications, because its non local properties improve capacity and speed information processing \cite{jozsa1, jozsa2, bennet1}. Control of entanglement is achievable in Ising model through of driven magnetic fields being introduced on the physical system. This is the case for teleportation. Different models of Ising interaction ($XX, XY, XYZ$ depending on interest of each author and physical systems being considered) are used to reproduce effects related with bipartite or multipartite systems \cite{berman1,wang2,aless1} and quantum dots \cite{brunner1, gau1}). 
  
Nowadays, quantum gate array computation is being experimentally explored to adapt it to stuff in which it can be settled, particularly in terms of noise control and reproduction of computational gates. It means, the interactions able to be considered to reproduce them \cite{cirac1,turchette1,shnirman1,chuang1,kane1,loss1,vrijen1}. Quantum dots and electronic gases are developments towards a scalable spin-based quantum computer, which can be controlled with electromagnetic interactions among neighboring spins to obtain universal quantum operations \cite{recher1, saraga1, kopp1} in terms of DiVincenzo criteria \cite{vincenzo1}. 

The aim of this paper is apply some non local properties recently reported by \cite{delgadoA} in the anisotropic Ising model and some control procedures on it \cite{delgadoB}, which naturally reproduces non local gates useful for teleportation. These gates reduce teleportation algorithm to a driven magnetic interaction based on matter to transfer spin quantum states. Several variants of this process are presented, thus as their extension to multiple qubits teleportation.

\section{Analytic evolution for anisotropic Ising model in three dimensions}
\begin{figure}[pb] 
\centerline{\psfig{width=23pc,file=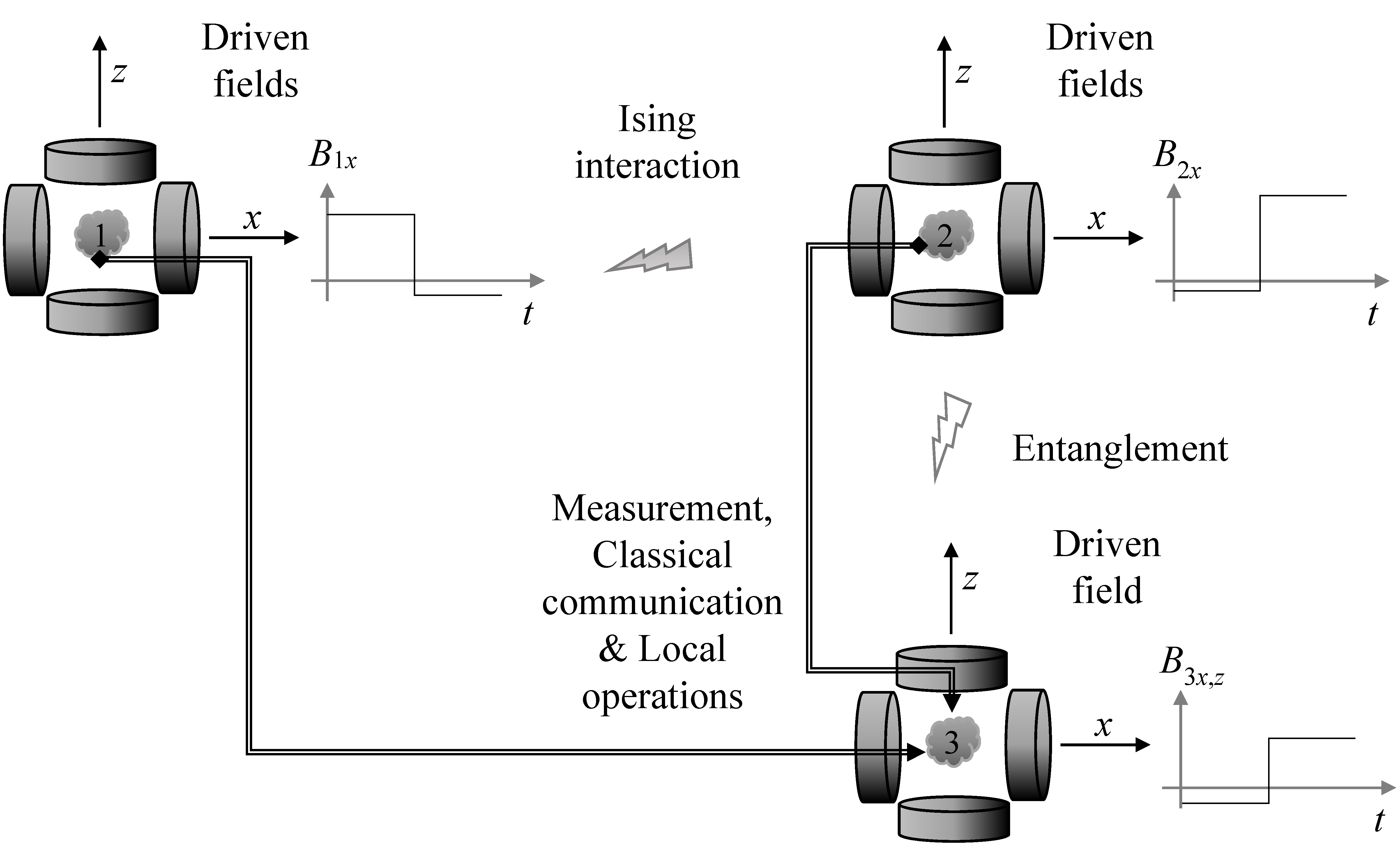}}
\vspace*{8pt}
\caption{Ising interaction between two qubits with driven local magnetic fields to produce teleportation on a third qubit entangled with the system.}
\label{fig1}
\end{figure}

The use of magnetic systems as quantum resources is a basis on which quantum applications could be settled. From quantum memories to quantum processors, matter susceptible of magnetic control is considered as stuff for quantum computation or quantum information processing. Particularly, Ising model, is a simple model of interaction  bringing an easy basis to generate and manipulate quantum states and entanglement particularly. In this model, as is shown in the upper part of Figure \ref{fig1}, two qubits interact via Ising interaction using additional local magnetic fields as driven elements. 

\subsection{Three dimensional anisotropic Ising model and notation}

Some results about anisotropic Ising model generalize their treatment and suggest an algebraic structure when it is depicted on Bell basis \cite{delgadoA}. Following this work, we focus on the following Hamiltonian for the bipartite anisotropic Ising model including an inhomogeneous magnetic field restricted to the $h$-direction ($h=1,2,3$ corresponding to $x,y,z$ respectively):

\begin{eqnarray} \label{hamiltonian}
H_h&=&-\mathbf{\sigma_1 \cdot J \cdot \sigma_2}+\mathbf{B_1 \cdot \sigma_1}+\mathbf{B_2 \cdot \sigma_2} \nonumber \\
&=& -\sum_{k=1}^3 J_k {\sigma_1}_k {\sigma_2}_k+{B_1}_h {\sigma_1}_h+{B_2}_h {\sigma_2}_h\ 
\end{eqnarray}

\noindent which includes several models considered in the previously cited works. Following the definitions and the notation adopted in \cite{delgadoA}, we introduce the scaled parameters:

\begin{eqnarray} \label{defs}
{b_h}_\pm &=& \frac{{B_h}_\pm}{{R_h}_\pm} , \quad {j_h}_\pm=\frac{{J_{\{h\}}}_\mp}{{R_h}_\pm} \quad \in [-1,1] \\ \nonumber \\ \nonumber 
{\rm with:} && J_{\{ h \}\pm} \equiv {J_{i,j}}_\pm=J_i \pm J_j \\ \nonumber 
&& B_{h \pm}=B_{1_h}\pm B_{2_h} \\ \nonumber
&& {R_h}_\pm = \sqrt{{B_h^2}_\pm+{J_{\{ h \}}^2}_\mp}
\end{eqnarray}

\noindent being $h,i,j$ a cyclic permutation of $1,2,3$ and simplified by $\{ h \} \equiv i,j$. Additionally, we introduce the same nomenclature introduced in \cite{delgadoA}based on different subscripts than those inherited by the computational basis: greek scripts for $-1,+1$ or $-,+$, for scripts in states and operators (meaning $-1,+1$ in mathematical expressions respectively); capital scripts $A, B, ...$ for $0, 1$ as in the computational basis; and latin scripts $h,i,j,k,...$ for spatial directions $x,y,z$ or $1,2,3$. Bell states become in last notation:

\begin{eqnarray} \label{bellnotation}
\left| \beta_{--} \right> \equiv \left| \beta_{00} \right> &,& 
\left| \beta_{-+} \right> \equiv \left| \beta_{01} \right> \\ \nonumber 
\left| \beta_{+-} \right> \equiv \left| \beta_{10} \right> &,&  
\left| \beta_{++} \right> \equiv \left| \beta_{11} \right>    
\end{eqnarray}

\subsection{Algebraic structure of evolution operator}

In agreement with last notation, energy levels in (\ref{hamiltonian}) are denoted by eigenvalues ${E_h}_{\mu \nu} ({E_h}_{--},{E_h}_{-+},{E_h}_{+-},{E_h}_{++})$ and they are:

\begin{eqnarray} \label{eigenvalues2}
{E_h}_{\mu \nu}&\equiv&\mu J_h+\nu {R_h}_{-\mu}  =\mu J_h+\nu \sqrt{{B_h}^2_{-\mu}+{J^2_{\{h\}}}_{\mu}}
\end{eqnarray}

\noindent with their eigenvectors reported in \cite{delgadoA}. As there, by introducing the following definitions:

\begin{eqnarray} \label{reduced}
{\Delta_h}_\mu^\nu &=& \frac{t}{2} ({E_h}_{\mu +}+\nu {E_h}_{\mu -})=
\left\{
\begin{array}{l l}
\mu J_h t & \rm{if} \quad \nu=+ \\
{R_h}_{-\mu} t & \rm{if} \quad \nu=- 
\end{array}
\right. \\
{e_h}_\alpha^\beta &=& \cos {\Delta_h}_\alpha^- + i \beta {j_h}_{-\alpha} \sin {\Delta_h}_\alpha^- \\ \nonumber
{d_h}_\alpha &=& {b_h}_{-\alpha} \sin {\Delta_h}_\alpha^-
\end{eqnarray}

\noindent the evolution operators in Bell basis are in matrix form:

\begin{eqnarray} \label{mathamiltonian1}
{U_{1}}(t)=& \left(
\begin{array}{c|c|c|c}
{e^{i {\Delta_1}_-^+}{e_1}_-^-}^* & i e^{i {\Delta_1}_-^+}{d_1}_-      & 0           & 0      \\
\hline
i e^{i {\Delta_1}_-^+}{d_1}_- & e^{i {\Delta_1}_-^+}{{e_1}_-^-}  & 0           & 0            \\
\hline
0         & 0          & {e^{i {\Delta_1}_+^+}{e_1}_+^+}^*  & -i e^{i {\Delta_1}_+^+}{d_1}_+  \\
\hline
0         & 0          & -i e^{i {\Delta_1}_+^+}{d_1}_+     & {e^{i {\Delta_1}_+^+}{e_1}_+^+} 
\end{array}
\right)  &\in {\bf S}^*_1 \\[3mm] \label{mathamiltonian2}
{U_{2}}(t)=& \left(
\begin{array}{c|c|c|c}
e^{i {\Delta_2}_+^+}{{e_2}_+^+}^*    &   0   &   0   & - e^{i {\Delta_2}_+^+}{d_2}_+  \\
\hline
0  &  e^{i {\Delta_2}_-^+}{{e_2}_-^+}^* &  -e^{i {\Delta_2}_-^+}{{d_2}_-}  & 0        \\
\hline
0  &  e^{i {\Delta_2}_-^+}{{d_2}_-} &  e^{i {\Delta_2}_-^+}{{e_2}_-^+}  & 0           \\
\hline
e^{i {\Delta_2}_+^+}{d_2}_+    &   0   &   0   & e^{i {\Delta_2}_+^+}{{e_2}_+^+}  
\end{array} 
\right) &\in {\bf S}^*_2 \\[3mm] \label{mathamiltonian3}
{U_{3}}(t)=& \left(
\begin{array}{c|c|c|c}
e^{i {\Delta_3}_-^+}{{e_3}_-^+}^* & 0 & i e^{i {\Delta_3}_-^+}{d_3}_-      & 0        \\
\hline
0  &  e^{i {\Delta_3}_+^+}{{e_3}_+^+}^* & 0 & i e^{i {\Delta_3}_+^+}{d_3}_+           \\
\hline
i e^{i {\Delta_3}_-^+}{d_3}_- & 0 &  e^{i {\Delta_3}_-^+}{{e_3}_-^+}      & 0         \\
\hline
0  &  i e^{i {\Delta_3}_+^+}{d_3}_+ & 0 & e^{i {\Delta_3}_+^+}{{e_3}_+^+} 
\end{array}
\right) &\in {\bf S}^*_3
\end{eqnarray}

\noindent where clearly $U_{h}(t)$ have a $2 \times 2$ block structure, having each one the semidirect product structure: $U(2)=U(1)\times SU(2)$. Thus, $U_{h}(t)$ belongs to subgroups: ${\bf S}^*_1, {\bf S}^*_2, {\bf S}^*_3 \subset SU(4)$ (defined in \cite{delgadoA}). These subgroups are characterized by their block structure and properties inherited from (\ref{mathamiltonian1}-\ref{mathamiltonian3}) when ${b_h}_{\pm \alpha}, {j_h}_{\pm \alpha}$ are fixed. Then, these subgroups have a semidirect product structure: ${S}^*_h=U(1) \times SU(2) \times SU(2)$. Thus, identity and inverses are included in each subgroup, while there are closure in the product. At same time, quantum states in ${\mathcal H}^{\otimes 2}$ become split in a direct sum of two subspaces generated each one by pairs of Bell states. This structure is essential in the current work because it assures the existence of solutions in their control.

\subsection{Block structure}

Blocks are elements of $U(2)$ as is reported in \cite{delgadoA,delgadoB}. Their general structure is:

\begin{eqnarray}\label{sector}
{s_h}_{j} &=& {e^{i {\Delta_h}_\alpha^+} \left(
\begin{array}{cc}
{{e_h}_\alpha^\beta}^* & -q i^h {d_h}_\alpha   \\
q {i^*}^h {d_h}_\alpha & {{e_h}_\alpha^\beta}    
\end{array}
\right)} \\
&{\rm with: }&
\alpha = (-1)^{h+j+1} \nonumber \\ 
&& \beta = (-1)^{j(h+l_j-k_j+1)} \nonumber \\ 
&& q = \beta (-1)^{h+1} \nonumber
\end{eqnarray}

\noindent being $h$ the spatial direction of magnetic field associated; $j=1, 2$ an ordering label for each block as it appears in the rows of the evolution matrix; $k_j, l_j$ are the labels for its rows in $U_h(t)$ (by example, in ${s_2}_1$, $k_2=2, l_2=3$ labels for the rows of second block, $j=2$, in $U_{h=2}(t)$). Note that $\det ({s_h}_{j})={e^{2 i {\Delta_h}_\alpha^+}}$ is unitary. Last structure lets introduce generation of control operations in terms of factorization of special unitary matrices in $SU(4)$ \cite{delgadoE}. 

\section{Controlled gates generated by anisotropic Ising interaction driven by magnetic fields}
As was reported in \cite{delgadoB}, controlled gates on each block could be constructed in two pulses:

\begin{equation}
U_h(T=t+t')=U_h(t')U_h(t)
\end{equation}

For that purpose, one block should be diagonalized into $I_2$ and another block antidiagonalized into $\sigma_1$ or $i \sigma_2$(which forms straight assumed wrote for the Bell basis instead for the computational basis). Prescriptions to achieve last forms on each block were obtained by \cite{delgadoB} and they are:

\begin{eqnarray} \label{presc1}
\frac{{B_h}_{\alpha}}{{J_{\{h\}}}_{-\alpha}}= \xi,
\frac{{B'_h}_\alpha}{{J'_{\{h\}}}_{-\alpha}}=-\xi^{-1},
\frac{{B_h}_{-\alpha}}{{J_{\{h\}}}_\alpha}= \chi,
\frac{{B'_h}_{-\alpha}}{{J'_{\{h\}}}_{\alpha}}=\chi& \\
\frac{|{J_{\{h\}}}_{-\alpha}| t}{(2n_{-\alpha}+1)} = \frac{|{J'_{\{h\}}}_{-\alpha}| t'}{(2n'_{-\alpha}+1)|\xi|}=\frac{\pi}{2\sqrt{\xi^2+1}} 
\quad \quad \nonumber 
\end{eqnarray}

\begin{eqnarray}\label{presc2}
&{\rm where:}& |\xi|= \frac{-AB\pm\sqrt{A^2+B^2-1}}{B^2-1} \\
&& \chi^2 = \left( \frac{2n_\alpha \sqrt{\xi^2+1}}{S_\alpha (2n_{-\alpha}+1)+P_\alpha S'_\alpha (2n'_{-\alpha}+1)|\xi|} \right)^2 -1  \nonumber \\ 
&& A=\frac{(2n_{-\alpha}+1)J_h}{2(m_\alpha+n_\alpha)|{J_{\{h\}}}_{-\alpha}|} \nonumber \\ 
&& B=\frac{(2n'_{-\alpha}+1)J'_{h}}{2(m_\alpha+n_\alpha)|{J'_{\{h\}}}_{-\alpha}|} \nonumber \\
&& P_\alpha={\rm sign} ({J'_{\{h\}}}_\alpha {J_{\{h\}}}_{ \alpha}) \nonumber \\ 
&& S_\alpha=|\frac{{J_{\{h\}}}_\alpha}{{J_{\{h\}}}_{-\alpha}}|, S'_\alpha=|\frac{{J'_{\{h\}}}_\alpha}{{J'_{\{h\}}}_{-\alpha}}| \nonumber
\end{eqnarray}

\noindent where $\alpha$ corresponds to diagonal block and $-\alpha$ to antidiagonal one. Also, $n_{\pm \alpha}, n'_{\alpha}, m_{\alpha} \in {\bf Z}$. These prescriptions let to obtain the design parameters $B_{\pm \alpha}, B'_{\pm \alpha}, t, t'$ in terms of strengths $J_{h},J'_{h},{J_{\{h\}}}_{\pm \alpha}, {J'_{\{h\}}}_{\pm \alpha}$. Note that parameters $|\xi|,\chi^2$ should fulfill positivity restrictions.

If ${s_h}^{\pm \alpha}_j, {s'_h}^{\pm \alpha}_j$ are the respective blocks $\alpha$ (diagonal) and $-\alpha$ (antidiagonal) for $U_h(t), U_h(t')$. Then, last prescriptions generate the combined blocks in $U_h(T)$:

\begin{eqnarray} \label{sectors1}
{s'_h}^{\alpha}_{j} {s_h}^{\alpha}_{j} &=& (-1)^{m_\alpha}I_2 \\ \label{sectors2}
{s'_h}^{-\alpha}_{j} {s_h}^{-\alpha}_{j} &=& 
\left(
\begin{array}{cc}
0 & 
(-1)^{s_\alpha}  \\
(-1)^{h+s_\alpha} & 
0    
\end{array}
\right) \nonumber \\
&=& (-1)^{s_{\alpha}}i^{h  {\rm mod}  2} \sigma_{1+h  {\rm mod}  2}
\end{eqnarray}

\noindent with $s_\alpha$ becoming an integer or a semi-integer in terms of the additional condition:

\begin{eqnarray}\label{diagadiagphase}
2(m_\alpha+n_\alpha)&=&-(h+{\rm sign} (q \beta b'_{h_{\alpha}}j_{h_{\alpha}}) + \quad 2(n_{-\alpha}+n'_{-\alpha}-s_{-\alpha}+1))
\end{eqnarray}

\noindent which shows that $s_\alpha$ is a semi-integer for $h=2$.

\section{Quantum teleportation based on Ising interaction}

\subsection{One qubit state teleportation using the ${\mathcal A}_{1,2}^{0,\frac{\pi}{2}}$ gate}

Driven Ising interaction generate generalized controlled gates presented in \cite{delgadoB}, ${\mathcal D}^{\phi}_h,{\mathcal A}_{h,j}^{\phi,{\varphi_h}_\alpha}$. First ones are near related with Evolution Loops combined with phase gates. Second ones are more like to $C^a NOT_b$ gates used in traditional gate array quantum computation. When $\phi=0,\varphi=\pi/2$, they are simply the matrices ${\mathcal A}_{h,j}^{0,{\frac{\pi}{2}}_\alpha}$ obtained from combination (in the structure of $U_h(t)$) of exactly each one of blocks in (\ref{sectors1}-\ref{sectors2}) with $m_\alpha, s_\alpha$ even:

\begin{eqnarray} \label{diagadiagforms}
{\mathcal A}_{1,1}^{0,\frac{\pi}{2}} = \left(
\begin{array}{c c c c}
0 & 1 & 0 & 0      \\
-1 & 0 & 0 & 0            \\
0 & 0 & 1 & 0  \\
0 & 0 & 0 & 1 
\end{array}
\right)  , 
{\mathcal A}_{2,1}^{0,\frac{\pi}{2}} = \left(
\begin{array}{c c c c}
0 & 0 & 0 & i      \\
0 & 1 & 0 & 0            \\
0 & 0 & 1 & 0  \\
i & 0 & 0 & 0 
\end{array}
\right) ,
{\mathcal A}_{3,1}^{0,\frac{\pi}{2}} = \left(
\begin{array}{c c c c}
1 & 0 & 0 & 0      \\
0 & 0 & 0 & 1            \\
0 & 0 & 1 & 0  \\
0 & -1 & 0 & 0 
\end{array}
\right) \\ \nonumber
{\mathcal A}_{1,2}^{0,\frac{\pi}{2}} = \left(
\begin{array}{c c c c}
1 & 0 & 0 & 0      \\
0 & 1 & 0 & 0            \\
0 & 0 & 0 & 1  \\
0 & 0 & -1 & 0 
\end{array}
\right),
{\mathcal A}_{2,2}^{0,\frac{\pi}{2}} = \left(
\begin{array}{c c c c}
1 & 0 & 0 & 0      \\
0 & 0 & i & 0            \\
0 & i & 0 & 0  \\
0 & 0 & 0 & 1 
\end{array}
\right), 
{\mathcal A}_{3,2}^{0,\frac{\pi}{2}} = \left(
\begin{array}{c c c c}
0 & 0 & 1 & 0      \\
0 & 1 & 0 & 0            \\
-1 & 0 & 0 & 0  \\
0 & 0 & 0 & 1 
\end{array}
\right)
\end{eqnarray}

Last suggests they can be used as natural operations replacing some standard gates if computing is done using non local quantum resources (remember that last expressions are based on non local Bell basis). One immediate application is the teleportation algorithm. We begin, as commonly, with the state $(\alpha \left| 0 \right>+\beta \left| 1 \right>) \otimes \left| \beta_{--} \right>$ with the first qubit to teleport and in possession of Alice. Bell state is first in possession of Bob, who send his first part to Alice. Then, Alice drives and Ising interaction as in our model, applying by example, ${\mathcal A}_{1,2}^{0,\frac{\pi}{2}}$ operation between the qubits in their current possession. With just this, they almost obtain the standard teleportation algorithm for one qubit \cite{bennet4}:

\begin{eqnarray}\label{teleportation}
(\alpha \left| 0 \right>+\beta \left| 1 \right>) \otimes \left| \beta_{00} \right> &=& \frac{\alpha}{2} \bigg( (\left| \beta_{00} \right>+\left| \beta_{10} \right>) \otimes \left| 0 \right> + 
(\left| \beta_{01} \right>+\left| \beta_{11} \right>) \otimes \left| 1 \right> \bigg)  \\
&+& \frac{\beta}{2} \bigg( (\left| \beta_{01} \right>-\left| \beta_{11} \right>) \otimes \left| 0 \right> + 
(\left| \beta_{00} \right>-\left| \beta_{10} \right>) \otimes \left| 1 \right> \bigg) \nonumber \\
\stackrel{{\mathcal A}_{1,2}^{0,\frac{\pi}{2}}}{\longrightarrow} & & \frac{\alpha}{2} \bigg( (\left| \beta_{00} \right>-\left| \beta_{11} \right>) \otimes \left| 0 \right> + 
(\left| \beta_{01} \right>+\left| \beta_{10} \right>) \otimes \left| 1 \right> \bigg) \nonumber \\
&+& \frac{\beta}{2} \bigg( (\left| \beta_{01} \right>-\left| \beta_{10} \right>) \otimes \left| 0 \right> + 
(\left| \beta_{00} \right>+\left| \beta_{11} \right>) \otimes \left| 1 \right> \bigg) \nonumber
\end{eqnarray}

\noindent where for simplicity, to identify easily the results in computational basis, we comeback at this point to the traditional notation for Bell states. Figure 2a and 2b show two alternatives of the algorithm using the gate ${{\mathcal A}_{1,2}^{0,\frac{\pi}{2}}}$, which is substituting the traditional $H_1 C^1NOT_2$ operation on computational basis.

\begin{figure}[pb] 
\centerline{\psfig{width=40pc,file=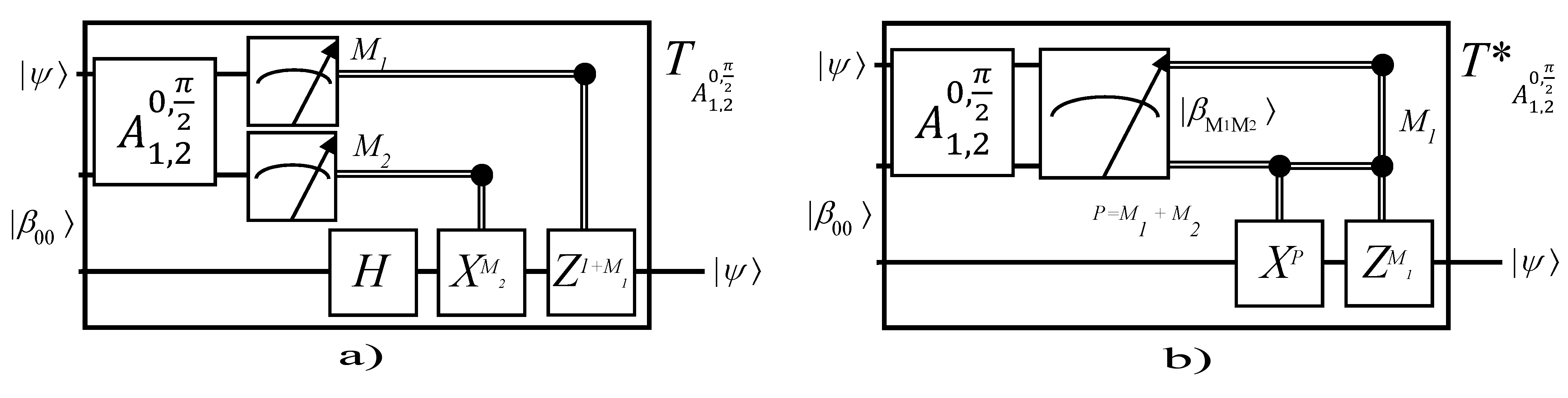}}
\vspace*{8pt}
\caption{Ising teleportation based on a) computational measurements, and b) non local measurements.}
\label{fig2}
\end{figure}

After to apply Ising interaction, Alice makes a measurement, by example in the computational basis on their qubits. Results are listed in Table \ref{tab1}, where $\left| - \right>, \left| + \right>$ are the eigenstates of $X$ (Teleportated state column in Table 1). Clearly a Hadamard gate is necessary at the end and in addition, if measurement result is $\left| M_1 M_2 \right>$, then it is required to apply $X^{M_2} Z^{1+M_1}$ to teleportate the original qubit 1 on qubit 3 ($M_1, M_2=0,1$ are in the traditional computational basis, not in the $+,-$ notation). Thus, Alice uses classical communication to send their outcomes to Bob, who finally applies the adequate gates (Complementary gates column in Table 1) to obtain exactly (until unitary factors) the original state $\alpha \left| 0 \right>+\beta \left| 1 \right>$, as is shown in Table 1 and Figure \ref{fig2}. 

There, $I,X,Y,Z,H$ are the traditional gates in computational basis, which should be applied on qubit 3 as is depicted in Table 1. We use these symbols to distinguish from $I_2, \sigma_1, \sigma_2, \sigma_3$ previously used (with exception of Hamiltonian (\ref{hamiltonian})), which are only matrix forms written for the Bell states corresponding to each subspace in the direct sum decomposition of ${\mathcal H}^{\otimes 2}$.

Another alternative for Alice and Bob is to make a measurement in Bell basis $\left| \beta_{\alpha \beta} \right>=\left| \beta_{M_1 M2} \right>$ (for both notations). In such case, Bob will need to apply $Z^{\frac{1+\alpha}{2}} X^{\frac{1-\alpha \beta}{2}}$ (with $\alpha, \beta=+,-$) or $Z^{M_1} X^{M_1 + M_2}$ (in the traditional notation $M_1, M_2=0,1$). Figure 3 shows this procedure. 
		
\begin{center}
\begin{table}[h]
\caption{\label{tab1} Measurement outcomes in two different basis, output state and complementary gates for teleportation algorithm based on ${\mathcal A}_{1,2}^{0,\frac{\pi}{2}}$.}
\centering
\begin{tabular}{l c c c}
\hline
Basis & Measurement & Teleportated state & Complementary gates \\
\hline
Computational &    $\left|00 \right>$ & $ \quad \alpha \left|+ \right>-\beta \left|- \right>$ & $Z_3 H_3$ \\
		& $\left|01 \right>$ & $ -\alpha \left|- \right>+\beta \left|+ \right>$ & $X_3 Z_3 H_3$ \\
		& $\left|10 \right>$ & $ \quad \alpha \left|+ \right>+\beta \left|- \right>$ & $H_3$ \\
		& $\left|11 \right>$ & $ \quad \alpha \left|- \right>+\beta \left|+ \right>$ & $X_3 H_3$ \\
		\hline
Non local & $\left|\beta_{--} \right>$ & $ \quad \alpha \left|0 \right>+\beta \left|1 \right>$ & $I_3$ \\
		& $\left|\beta_{-+} \right>$ & $ \quad \alpha \left|1 \right>+\beta \left|0 \right>$ & $X_3$ \\
		& $\left|\beta_{+-} \right>$ & $ \quad \alpha \left|1 \right>-\beta \left|0 \right>$ & $Z_3 X_3$ \\
		& $\left|\beta_{++} \right>$ & $ -\alpha \left|0 \right>+\beta \left|1 \right>$ & $Z_3$ \\
\hline
\end{tabular}
\end{table}
\end{center}

\subsection{General teleportation with other Ising gates and multiqubit states}

Last example has shown that gate ${\mathcal A}_{1,2}^{0,\frac{\pi}{2}}$ reproduce the traditional teleportation algorithm. Nevertheless, it can be achieved with any ${\mathcal A}_{h,j}^{0,\frac{\pi}{2}}$ gate. Following the steps in the last subsection, it is possible show that with any of these gates (for arbitrary $h=1,2,3$ and $j=1,2$) and inclusively if initial Bell state used is $\left| \beta_{AB} \right>$ instead of $\left| \beta_{--} \right>=\left| \beta_{00} \right>$ as before, teleportation procedure can be achieved if we use the following complementary gates (assuming non local measurements based on Bell states, $\left| \beta_{M_1 M_2} \right>$):

\begin{eqnarray}
&Z^{a^{M_1,M_2}_{h,j;A,B}}& X^{b^{M_1,M_2}_{h,j;A,B}} \\
&{\rm with:}& a^{M_1,M_2}_{h,j}=A+\frac{1}{8}\left(4(h-\frac{5}{2})^2-9 \right)(j-2-M_2)+(h-2)^2M_1 \nonumber \\
&& b^{M_1,M_2}_{h,j}=B+\frac{1}{8}\left(4(h-\frac{3}{2})^2-9 \right)(j-2-M_1)+(h-2)^2M_2
\end{eqnarray} 

The generalization of last procedure for a multiqubit state is easy. In the Figures 2 and 3, we define the procedure $T$ as a map from ${\mathcal H}^{\otimes 3}$ on ${\mathcal H}$ (here, ${\mathcal H}$ is a two level Hilbert space):

\begin{eqnarray}
&& T : {\mathcal H}_1 \otimes ({\mathcal H}^{\otimes 2})_{2,3} \rightarrow {\mathcal H}_3 \\
&& T((\alpha\left| \psi_\alpha \right>_1+\beta\left| \psi_\beta \right>_1) \otimes \left| \phi \right>_{2,3})=\alpha\left| \psi_\alpha \right>_3+\beta\left| \psi_\beta \right>_3 \nonumber
\end{eqnarray}

\noindent we can note that it is linear in the original state:

\begin{eqnarray}
T((\alpha\left| \psi_\alpha \right>_1+\beta\left| \psi_\beta \right>_1) \otimes \left| \phi \right>_{2,3}) = \alpha T(\left| \psi_\alpha \right>_1 \otimes \left| \phi \right>_{2,3}) + \beta T(\left| \psi_\beta \right>_1 \otimes \left| \phi \right>_{2,3})
\end{eqnarray}

\noindent While, state $\alpha\left| \psi_\alpha \right>_1+\beta\left| \psi_\beta \right>_1$ is separable from the remaining state $\left| \phi \right>_{2,3}$. In the last expressions, the subscripts are the labels for each one of the three qubits involved. Then, generalization is obvious. If we have the $n$-qubit state:

\begin{eqnarray}
\left| \psi \right>_{1,...,n} = \sum^1_{i_1,...,i_n=0} \alpha_{\{i_1,...,i_n\}} \bigotimes^n_{j=1} \left| i_j \right>_{j}
\end{eqnarray}

\noindent to teleportate, then, if we use the following state for the ancilla qubits:

\begin{eqnarray}
\left| \phi \right>_{n+1,...,2n} = \bigotimes^n_{j=1} \left| \beta_{A_j B_j} \right>_{n-1+2j,n-2+2j}
\end{eqnarray}

\noindent and we apply the teleportation protocol $T_j$ on qubits $j,n-1+2j,n-2+2j$ (here, each $T_j$ could be ${T_j}_{{\mathcal A}_{1,2}^{0,\frac{\pi}{2}}}$ or ${T^*_j}_{{\mathcal A}_{1,2}^{0,\frac{\pi}{2}}}$):

\begin{eqnarray}\label{multi}
\bigotimes^n_j T_j(\left| \psi \right>_{1,...,n} \otimes \left| \phi \right>)_{n+1,...,2n} &=& \sum^1_{i_1,...,i_n=0} \alpha_{\{i_1,...,i_n\}} \bigotimes^n_{j=1} T_j(\left| i_j \right>_{j} \otimes \left| \beta_{A_j B_j} \right>_{n-1+2j,n-2+2j}) \\
&=& \sum^1_{i_1,...,i_n=0} \alpha_{\{i_1,...,i_n\}} \bigotimes^n_{j=1} \left| i_j \right>_{n-2+2j}) = \left| \psi \right>_{n+2,n+4,...,2n-2,2n} \nonumber
\end{eqnarray}

Note in particular that not all single qubit teleportation algorithms can be the same type as well as the Bell state $\left| \beta_{A_j B_j} \right>$ being used for each one. Equal signs used in (\ref{multi}) are until unitary factors. These results show that controlled Ising model reproduces equivalent standard gates proposed in quantum teleportation.

\section{Conclusions}

Optics has been partially a dominant arena to developments in quantum information and quantum computation. Nevertheless, quantum storage and massive processing of information are tasks abler to develop on matter based quantum systems. Systems based on trapped ions, e-Helium, nuclear magnetic resonance, superconductors, doped silicon and quantum dots are developments on this trend, showing possibilities to set up stable and efficient implementations of those technologies \cite{klo1}. Thus, spin-based quantum computing has been developed in through experimental implementations using magnetic systems mainly and exploiting Ising interactions with different approaches \cite{john1}, together with control on quantum states, in particular  entanglement control. 

The procedure analyzed in this paper shows how Ising interaction could be controlled to generate gates able to produce teleportation. Because its linearity, these scheme shows the same generalization that traditional algorithm \cite{bennet4}. In addition, the different paths provided in the analysis, in terms of the Bell states used as resources and the field direction stated to generate the controlled gates, open other future research directions. One is the analysis of control in the process and its fidelity in terms of control parameters (interaction and fields strengths, time intervals). One more is about flexibility, as was stated in \cite{delgadoD}, in terms of viability to use resources with discriminate them, it means, to achieve teleportation independently the knowledge about non local resource being used. Finally, extensions to optimize the economy in terms of non local resources used in the process is open as in the general teleportation trend.

\end{document}